# On the Lyman $\alpha$ Emission of Starburst Galaxies

David Valls–Gabaud[1,2,3,4]

## ABSTRACT

Nearby starburst galaxies have consistently shown anomalous Ly$\alpha$/H$\beta$ ratios. By re-analysing the published IUE/optical observations, we show that most starbursts present a *normal* Ly$\alpha$ emission, consistent with case B recombination theory, provided extinction laws appropriate to their metallicities are used. This implies that extinction is more important than multiple resonant scattering effects. The anomalous emission and absorption lines present in a few remaining galaxies are simply explained if they are observed in the post-burst phase, between about $10^7$ and $10^8$ yrs after the start of the burst. We use updated stellar population synthesis models to show that anomalous ratios are produced by the aging of stellar populations, since the underlying stellar Ly$\alpha$ line is important in the cooler massive stars. The inferred low-duty cycle of massive star formation accounts naturally for the failure to detect large numbers of Ly$\alpha$–emitting galaxies in deep surveys and at high redshift. Some testable predictions of the proposed scenario are also discussed.

*Subject headings:* cosmology – galaxies: evolution – galaxies: formation – galaxies: stellar content – galaxies: interstellar matter –ultraviolet: galaxies

---

[1] Institut d'Astrophysique de Paris, 98 bis Bld Arago, F-75014 Paris, France.
[2] Institute of Astronomy, Madingley Road, Cambridge CB3 0HA, UK.
[3] Present address: Physics Department, Queen's University, Kingston, Ontario K7L 3N6, Canada.
[4] NSERC International Fellow.

# 1 INTRODUCTION

In any galaxy formation scenario young galaxies are predicted to present a strong Ly$\alpha$ emission line, essentially associated with the cooling of the gas and the subsequent formation of stars. The details of the emission depend on the assumed cooling and star formation rates during the collapse and infall of gas (Partridge & Peebles 1967, Kaufman 1976, Meier 1976, Rees 1988). Recent simulations of an inhomogeneous dissipational collapse (Baron & White 1987) also indicate a strong Ly$\alpha$ line.

Yet despite the large number of observations aimed at detecting such an emission, no strong Ly$\alpha$ sources have been found at high redshift that can be classified as protogalaxies (Davis 1980, Koo 1986, Pritchet & Hartwick 1987, Lowenthal et al. 1990, De Propis et al. 1993); both distant quasars and radio-galaxies have a Ly$\alpha$ emission probably dominated by non-thermal processes. A possible exception are the recently observed Ly$\alpha$ emission lines associated with damped Ly$\alpha$ systems (Hunstead, Pettini & Fletcher 1990, Lowenthal et al. 1991, Wolfe et al. 1992, Møller & Warren 1993).

One way to understand the conditions of Ly$\alpha$ emission at high redshift is to study the properties of nearby starburst galaxies. Several observations have consistently shown that their Ly$\alpha$ emission is quite peculiar, being very weak, absent or even in absorption (Meier & Terlevich 1981 [MT], Hartmann et al. 1984 [H84], Deharveng, Joubert & Kunth 1985 [DJK], Hartmann et al. 1988 [H88]). It was thus argued that their corresponding Ly$\alpha$/H$\beta$ ratios are inconsistent with case B recombination, rendering difficult the interpretation of their spectra. The current explanation is that the multiple resonant scattering with H I atoms increases the path length of Ly$\alpha$ photons, and accordingly the probability of absorption by dust (Hummer & Kunasz 1980). For this to hold, however, the optical depth of dust at Ly$\alpha$ must be relatively large and an homogeneous H I envelope with unit covering factor must surround the sources. In this paper we show that this is not the case, and we suggest an alternative mechanism.

# 2 THE LY$\alpha$/H$\beta$ RATIO IN STARBURSTS

We consider the entire sample of nearby starburst galaxies with IUE and optical observations (MT,H84,DJK, H88, Margon et al. 1988, Calzetti & Kinney 1992, Terlevich et al. 1993 [T93]). In order to disentangle extinction corrections from other effects, we study the dependence of the *extinction-corrected* Ly$\alpha$/H$\beta$ emissivity ratio on the *observed* (uncorrected) H$\alpha$/H$\beta$ flux ratio (Figure 1). The horizontal axis in this diagnostic diagram is thus a function of the optical depth of dust, while the vertical axis measures the deviation from standard recombination theory. If resonant scattering effects are unim-portant, galaxies should lie between the dotted lines in Figure 1. To derive the corrected Ly$\alpha$/H$\beta$ ratio the extinction law appropriate to the metallicity of the galaxy must be used. We adopt characteristic mean extinction laws corresponding to Galactic, LMC and SMC metallicities (Seaton 1979, Fitzpatrick 1986, Bouchet et al. 1985) to correct the ratios of each galaxy.

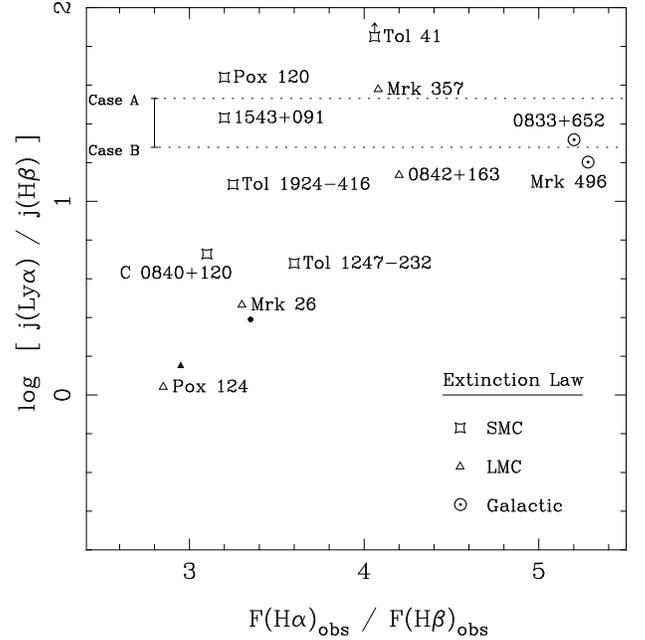

Figure 1: Ratio of *de-reddened* Ly$\alpha$/H$\beta$ emissivities as a function of the *observed* H$\alpha$/H$\beta$ flux ratio. The extinction law appropriate to the metallicity of each galaxy is indicated. The filled triangle and circle represent the positions of Mrk 357 and Mrk 496, respectively, at small apertures. Measured with a large (IUE) aperture, their positions (open symbols) agree with standard recombination theory.

As Figure 1 shows, most of the galaxies do indeed lie close to the range predicted by standard recombination theory, even when the dust optical depth is large (large Balmer decrement) and regardless of the H I column density present in each galaxy. Thus most galaxies have a *normal* Ly$\alpha$ emission and their Ly$\alpha$/H$\beta$ ratios agree with case B recombination, provided appropriate extinction laws are used. This in turn implies that resonant scattering effects are negligible, a fact that can easily be interpreted within the framework of inhomogeneous, multiphase models for the interstellar medium of starbursts (Neufeld 1991, Valls–Gabaud 1993 [VG93]) in which Ly$\alpha$ photons escape through the hot, ionised phase. The anticorrelation between the Ly$\alpha$/H$\beta$ ratio and the oxygen abundance that was found previously (DJK,MT,H88,T93) is not confirmed by our analysis, since galaxies in the entire range of metallicities, from Galactic (0833+652, Mrk 496) to SMC (Pox 120, Tol 41), present *normal* Ly$\alpha$/H$\beta$ ratios.

Some galaxies in the upper part of Figure 1 might have been overcorrected, in particular Tol 41, but this may be due to several effects. For instance the Balmer decrement can overestimate the extinction (as measured by the thermal radio continuum, e.g. Lequeux et al. 1981). Furthermore the extinction in the UV might be lower than in the visible (Fanelli et al. 1988). We expect that the high energy density contained in the radiation field of the starburst environment efficiently destroys the PAHs, as observed for example in the LMC (Sauvage, Thuan & Vigroux 1990). The extinction at Ly$\alpha$ must then decrease since PAHs make up an important part of the UV extinction (Désert, Boulanger & Puget 1990). Also, since the oxygen abundance is enhanced during a starburst an overestimation of the actual metallicity is produced (Gilmore & Wyse 1991) and may change the correction too.

A third of the sample lies in the lower part of Figure 1, and these galaxies (Pox 124, Mrk 26, Tol 1247-232 and C0840+120) are truly anomalous, since both their dust optical depth and their Ly$\alpha$ emission are small. Several effects can explain the discrepancy. The most important one is possibly due to the different apertures used for the IUE spectra and the optical ones. Calzetti & Kinney (1992) have reobserved in the optical three galaxies using the same (large) aperture as IUE, and found that the Balmer decrements were significantly larger than those measured using standard small apertures. In Figure 1 the positions of Mrk 357 (filled triangle) and Mrk 496 (filled circle), as derived from spectra taken with small apertures are indicated. Small apertures seem to underestimate the integrated Balmer decrement, and accordingly the Ly$\alpha$/H$\beta$ ratio, so it seems that aperture effects may account for the apparent anomalous ratios. Another possibility is of course the destruction by enhanced scattering in a uniform H I envelope and absorption by dust. Yet why this process should be efficient in anomalous galaxies but not in normal ones is unclear. Moreover, this effect would require an improbably large H I column density of $10^{23}$ cm$^{-2}$ for Pox 124.

Even if these effects may account for the remaining anomalous galaxies, neither of them explains the presence of Ly$\alpha$ lines in *absorption* detected in Mrk 309, Mrk 347 and Mrk 499 (H88). Another mechanism must necessarily play an important role. We suggest below that this is the aging of the stellar populations.

## 3 LY$\alpha$ EQUIVALENT WIDTHS AND AGING

In principle, if an homogeneous H I layer covers most of the H II regions, the equivalent width (EW) of the Ly$\alpha$ line in a galaxy spectrum is difficult to interpret, since line photons will be affected by multiple resonant scatterings, while continuum photons will not. However, this effect is negligible in nearby starbursts (§2), and thus the Ly$\alpha$ EW can be used to infer the properties of the underlying stellar populations.

An important and previously unaccounted-for effect which strongly reduces the Ly$\alpha$ emission is the fact that late B stars have very large Ly$\alpha$ equivalent widths in absorption, as Figure 2 shows (Valls–Gabaud 1991). The earlier B and O stars ($T_{\rm eff} \geq 22,000$ K) have small EWs, but the later ones show an exponential increase with decreasing temperature. This trend is clearly confirmed by OAO-2 (Savage & Panek 1974) and *Copernicus* observations (Vader et al. 1977), in addition to being consistent with predictions of NLTE models (Fig. 2). In the 30,000 K $\leq T_{\rm eff} \leq 10,000$ K range, the trend for main-sequence stars is

$$W_*(\text{Ly}\alpha) \approx -420 \text{ Å } \exp\left(-\frac{T_{\rm eff}}{6,100 \text{ K}}\right), \qquad (1)$$

where *absorption* EW is denoted with a minus sign.

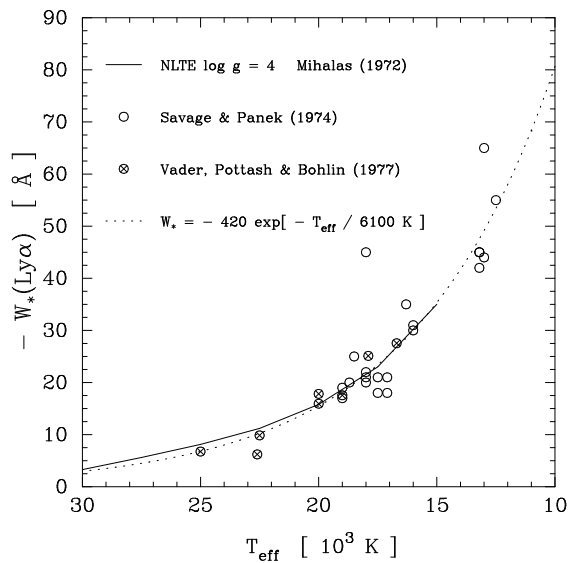

Figure 2: Equivalent width of the stellar Ly$\alpha$ absorption line as a function of effective temperature. The continuous line gives the theoretical predictions from NLTE models, while the dotted line is an empirical trend (Eq. 1).

No observations were found for early A type stars, except for Vega which has about 100 Å (Praderie 1981) and confirms the trend given by Eq. (1). Although lower temperature (F, G and M) stars may have a Lyα *emission* produced by the chromosphere (see e.g. Landsman & Simon 1993), their contribution to the integrated UV flux is negligible for ages less than $10^8$ yrs. Massive OB supergiants might develop a P Cygni profile in the Lyα line with a net EW in emission of less than 1 Å (Mihalas & Hummer 1974) although the details depend strongly on the wind properties (Sellmaier et al. 1993). In any case, this shorter lived phase does not affect the predictions for integrated spectra for ages less than $10^8$ yrs which will be dominated (at Lyα) by main sequence stars.

We have included the effect of the stellar Lyα line in the updated stellar population synthesis code of Bruzual and Charlot (1993) to estimate the net EW of the actual line. Figure 3a shows the evolution of the total EW ($W_{tot}(Ly\alpha)$) as a function of time for a burst and a constant star formation rate. The results are very robust and do not depend strongly on the slope $\alpha$ of the initial mass function (IMF), nor on its upper cutoff $m_{up}$, except in the very early phases ($t \leq 10^6$ yrs). During the first $10^{6.5}$ yrs, the EW is about 250 Å, decreasing sharply afterwards so that by $10^7$ yrs the Lyα line becomes negligible or indeed is in absorption. The time scale corresponds to the lifetime of the later B stars which are the only significant contributors to the far UV at that time. Note that the strong increase after $10^9$ yrs is produced by the emission from planetary nebulae, which is irrelevant for the galaxies studied here, but may be important in other galaxies (Hansen et al. 1991).

Figure 3b shows the large contribution of the underlying stellar population to the *absorption* EW ($W_{abs}$), typically 50 Å at $10^8$ yrs in the burst case. In the constant star formation rate case it is much less important since massive OB stars are continuously being formed and dominate the UV spectrum, and their integrated $W_{abs}$ ($\sim$−10 Å) is always much smaller than the emission one. The use of low metallicity stellar evolutionary tracks and spectra (Olofsson 1989, Mas-Hesse & Kunth 1991) does not change these values significantly.

In order to test whether these results are actually observed, we plot in Figure 4 the net Lyα EW versus the (U−B) colour for the sample of nearby starbursts (crossed circles). Galaxies with no published (U−B) colours were placed at the mean and median colour of the sample, (U−B)$_m$ = −0.4 (open circles) and are likely to be in the range −0.3 to −0.6. While the constant SFR is ruled out, given the distribution of galaxies in this diagram, the burst case provides an excellent fit and indicates several interesting properties:

*i)* in the *same* range of very blue colours there is a continuous distribution of galaxies, ranging from strong Lyα emitters to galaxies showing Lyα in absorption.

*ii)* the main parameter of this distribution is the age, since for populations younger than about $10^7$ yrs the Lyα emission is normal and the EW is large, while for older bursts the Lyα line is smaller or becomes in absorption, independently of the details of the IMF.

The aging of the stellar populations in a burst is thus a simple unifying mechanism that naturally accounts for the presence of normal Lyα emitters, galaxies with 'anomalous' Lyα/Hβ ratios, and Lyα lines observed in absorption. If a galaxy is observed in the 'active' burst phase, normal Lyα/Hβ ratios are observed. This is the case for most of the galaxies in the sample. Since the lines are unresolved, the measured flux scales as $(1-|W_{abs}|/W_{tot})$ and decreases with time (Fig. 3) so that seemingly 'anomalous' Lyα/Hβ ratios appear if the galaxy is observed in the 'post-burst' phase, about $10^7$ yrs after the start of the burst. When the burst is observed at an older age, between a few $10^7$ and $10^8$ yrs, the Lyα line is in absorption (Fig. 4).

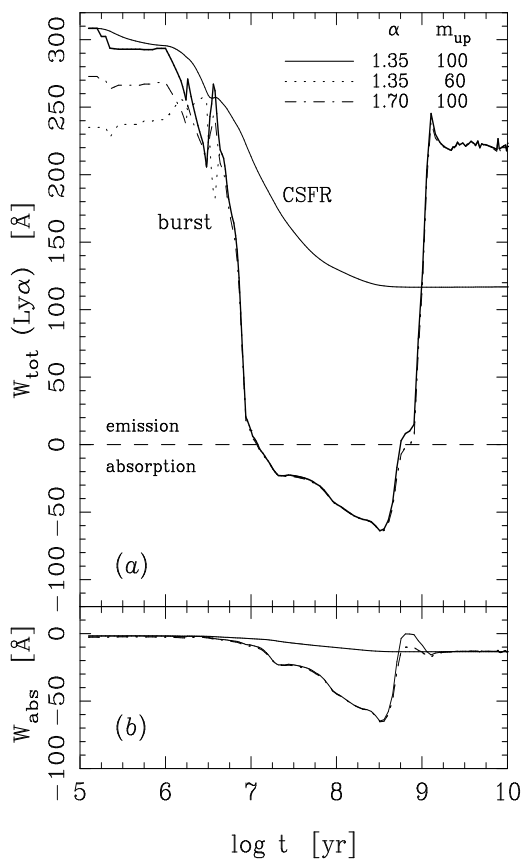

Figure 3: Evolution of the Lyα equivalent width for a burst with the three indicated IMFs and a constant star formation rate ($m_{low} = 0.1 M_\odot$ in all cases). *(a)* Total equivalent width (emission+absorption). After $\sim 6 \cdot 10^6$ yrs, the EW is independent of the IMF. *(b)* Integrated absorption equivalent width for the same cases.

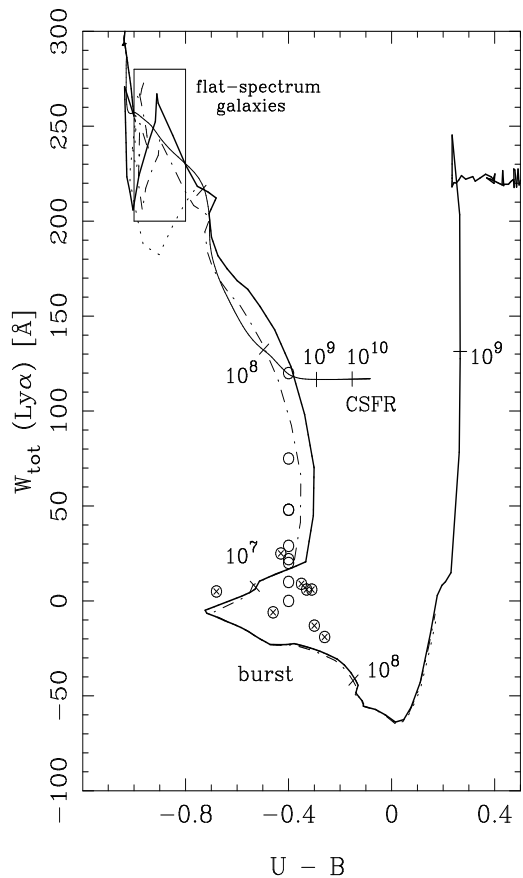

Figure 4: Lyα equivalent width as a function of the $U-B$ colour, for the same star formation rates and IMFs indicated in Fig. 3. Observations (crossed circles) and estimated uncertainties are marked. Open circles represent galaxies with measured Lyα EWs, but with no published colours (see text). The predicted position of flat spectrum galaxies is also indicated.

The Lyα line is far more sensitive to the aging of the stellar populations than the Balmer or the [OII]λ3727, [OIII]λλ4959,5007 lines since the stellar absorption line is much larger for Lyα than for the latter lines. Therefore, unlike the star formation rate obtained through the Hα emission (Kennicutt 1983), for which the stellar absorption line is always relatively small (e.g. Díaz 1988), a star formation rate derived from the Lyα line is highly unreliable, essentially because SFR(Lyα) ∝ EW(Lyα) × SFR(UVcont). A rate of star formation derived from the UV continuum flux or Balmer emission lines is more suitable since it is less biased. A contribution by supernovae remnants or shocks is not required to explain the Lyα line in these starbursts.

## 4 IMPLICATIONS FOR HIGH REDSHIFT GALAXIES

Ultra-deep imaging surveys (eg. Tyson 1988, Cowie et al. 1988) have revealed a population of possible 'flat-spectrum' galaxies. Our proposed scenario predicts that for those galaxies with flat rest-frame spectra (U−B∼−0.9), a conspicuous Lyα emission should be present with large EWs of the order of 250 Å (see Figure 4). Yet we expect that only a small fraction will be visible due to the short lived nature of the flat spectrum phase in U−B. Furthermore if high redshift radio galaxies have similar spectra (eg. Lilly 1989), the maximum EW expected from the associated young stellar population cannot exceed 300 Å, which is close to the observed values, although a non-thermal contribution by the nucleus or by shocks cannot be excluded.

In addition, this scenario easily explains the lack of success of deep imaging surveys in detecting primordial galaxies through their Lyα emission. Since any SFR can be decomposed into a series of short bursts (e.g. Searle, Sargent & Bagnuolo 1973), the Lyα emission from a galaxy is a strongly varying function of time. If $N_b$ is the number of bursts per galaxy in a Hubble time, and $t_b$ the duration of the burst in Lyα, the fraction $f_b$ of galaxies that are observable in Lyα emission is

$$f_b = N_b t_b / t_H \sim 10^{-3} h N_b \ . \qquad (2)$$

The number of bursts $N_b$ must satisfy the constraint that the total mass in stars formed in a Hubble time is the same as that produced by the average SFR in that period, hence

$$N_b \approx 50 h^{-1} \left(\frac{\mathrm{SFR}_b/<\mathrm{SFR}>}{20}\right)^{-1} \left(\frac{t_b}{10^7 \mathrm{\ yrs}}\right)^{-1} \qquad (3)$$

Then Eq. (2) implies that in the deep volumes sampled by the present surveys, only about 5% of the galaxies will be visible in Lyα. The constraints set by Lyα surveys (Pritchet & Hartwick 1990, Lowenthal et al. 1990, De Propis et al. 1993) on the theoretical scenarios of galaxy evolution are therefore much less stringent, by at least one order of magnitude, on the number density of Lyα emitters at a given flux. We also note that this argument explains the low rate of Lyα emission detections in damped Lyα systems (VG93). Ongoing imaging (Thompson, Djorgovski & Trauger 1992) and spectral surveys (Cowie et al. 1988) can provide a further test of this mechanism by setting limits on the duty cycle.

## 5 DISCUSSION AND CONCLUSIONS

A simple test of the scenario is provided by observations in Lyα and in the K band since the excursion to the red before $10^7$ yrs in Figure 4 is produced by red supergiants (RSG). In fact the most 'active' starbursts are known to have a K magnitude dominated by RSG (Devereux 1989), so they must also present Lyα emission. This is true even if they contain large amounts of dust, since most (>70%) of the galaxies analysed here have been detected by IRAS and have large FIR luminosities. Lyα photons can escape from such galaxies precisely because the starburst has established an inhomogeneous, multiphase interstellar medium (VG93).

Another interesting test is provided by Wolf-Rayet galaxies (eg. Conti 1991). Mrk 309 seems to present the 4650 Å WR bump (H88), yet its Lyα line is in absorption. If the spectra are confirmed, our scenario predicts that the WR feature is produced by the least massive stars that can evolve into the WR stage.

Should our model be correct, broad-band photometry of the recently detected high $z$ radio-quiet galaxies (Lowenthal et al. 1991, Steidel et al. 1991, Maccheto et al. 1993) will reproduce the pattern indicated in Fig. 4. For example we predict for object G2 at $z = 3.428$ (Maccheto et al. 1993) a rest-frame U−B≈−0.6 or bluer if EW(Lyα)≥163 Å, and similarly for the other galaxies.

In summary, the re-analysis of the sample of nearby starburst galaxies, combined with updated stellar population synthesis has shown that:

1) Most starbursts present normal, case B recombination Lyα emission, and extinction is more important than multiple resonant scattering effects. There is no anticorrelation between Lyα emission and metallicity.

2) The underlying stellar Lyα line is important and must be taken into account in any population synthesis study of massive stars.

3) The few galaxies with anomalous Lyα/Hβ ratios and absorption Lyα lines can be interpreted as galaxies observed in the post-burst phase, between about $10^7$ and $10^8$ yrs.

4) The EW(Lyα) is a good indicator of the burst age and does not depend strongly on the details of the IMF. The star formation rate derived from the Lyα luminosity is highly unreliable.

5) Deep Lyα imaging surveys can only detect about 5% of the total number of galaxies, and place less restrictive constraints on galaxy formation scenarios than previously thought.

I would like to thank M. Bithell, S. Charlot, C. Pritchet, M. Pettini and J. Lowenthal for interesting discussions.